\documentclass[journal]{IEEEtran}

\ifCLASSINFOpdf
\else
   \usepackage[dvips]{graphicx}
\fi
\usepackage{url}
\usepackage{graphicx}
\usepackage{amsmath} 
\usepackage{amssymb}
\begin{document}

% \title{Multi-Stage Speech Enhancement via Alternating Approach and Putt Models}

\title{Alternating Approach-Putt Models for Multi-Stage Speech Enhancement}
\author{Iksoon Jeong, Kyung-Joong Kim, and Kang-Hun Ahn
\thanks{This research was supported by Basic Science Research Program through the National Research Foundation of Korea(NRF) funded by the Ministry of Education(RS-2023-00246572).}
\thanks{Iksoon Jeong and Kyung-Joong Kim are with Department of Physics, Chungnam National University, Daejeon 34134, Republic of Korea (email: jeongis0203@gmail.com).}
\thanks{Kang-Hun Ahn is with Department of Physics, Chungnam National University, Daejeon 34134, Republic of Korea, and with Hearing Loss Research Lab., Deep Hearing Corp., Republic of Korea(email:ahnkanghun@gmail.com )}}

% \markboth{Journal of \LaTeX\ Class Files, Vol. 14, No. 8, August 2015}
% {Shell \MakeLowercase{\textit{et al.}}: Bare Demo of IEEEtran.cls for IEEE Journals}
\maketitle

\begin{abstract}
Speech enhancement using artificial neural networks aims to remove noise from noisy speech signals while preserving the speech content. However, speech enhancement networks often introduce distortions to the speech signal, referred to as artifacts, which can degrade audio quality. In this work, we propose a post-processing neural network designed to mitigate artifacts introduced by speech enhancement models. Inspired by the analogy of making a `Putt' after an `Approach' in golf, we name our model PuttNet. 
We demonstrate that alternating between a speech enhancement model and the proposed Putt model leads to improved speech quality, as measured by perceptual quality scores (PESQ), objective intelligibility (STOI), and background noise intrusiveness (CBAK) scores. Furthermore, we illustrate with graphical analysis why this alternating Approach outperforms repeated application of either model alone.
\end{abstract}


\IEEEpeerreviewmaketitle

\section{Introduction}

\IEEEPARstart{R}{ecent}  advances in deep learning have significantly improved speech enhancement (SE) systems in suppressing background noise. However, most single-stage approaches struggle with a fundamental trade-off between aggressive noise suppression and speech distortion. 
 This phenomenon is likely due to artifacts resulting from the neural networks designed to suppress noise, which may inadvertently damage some of the language-relevant components of the audio signal\cite{Iwamoto1, Iwamoto2}.

To address these challenges, multi-stage speech enhancement has emerged as a promising strategy. By decomposing the enhancement process into multiple stages—typically including artifact correction and phase refinement—these models enable progressive refinement and better separation of tasks.
For example, Wang and Wang\cite{Wang} utilized a diffusion model\cite{Song,Ho} conditioned on a speech enhancement network, demonstrating enhanced robustness under extremely noisy environments. Similarly, Lemercier et al. showed that integrating a diffusion model following a predictive framework yields significant improvements in both speech enhancement and dereverberation\cite{Lemercier}.

\begin{figure}[htb]
 \centering
  \centerline{\includegraphics[width=\columnwidth ]{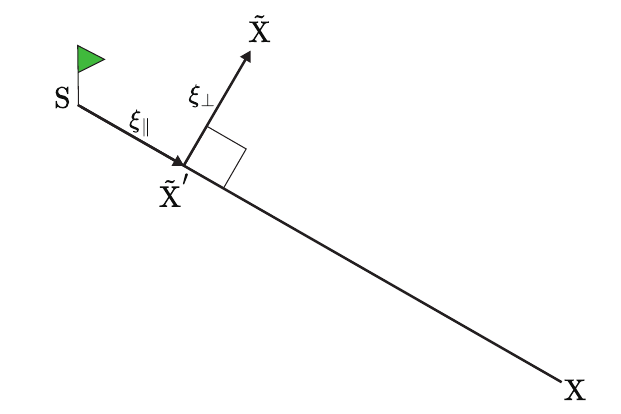}}

\caption{A schematic showing the definition of the artifact vector \boldmath$\xi_\perp$. The artifact vector \boldmath$\xi_\perp$ represents the foot of the perpendicular dropped from 
$\tilde{\bf X}$
 onto the line connecting clean speech ${\bf S}$
 and noisy sound ${\bf X = S+N}$
 (where 
${\bf N}$ denotes noise).}
\label{Schematic}

\end{figure}
Both of the aforementioned studies share a common Approach: employing a predictive model first, followed by a generative model. Predictive models, trained via supervised learning, produce outputs based on given inputs but are susceptible to underfitting or overfitting. In both studies, a diffusion model was utilized as the generative component. 
Diffusion models, however, have the drawback of being computationally intensive. 

%The reason why the denoising diffusion process is slow and it involves an iterative procedure and inherently operates as a sequential process.  

% \begin{figure}[htb]
%     \centerline{\includegraphics[width=\columnwidth]{figs/s1.pdf}}
%     \caption{An example of a result from our Approach-and-Putt model. The text written in each figure is a result of an STT.
%     (a) The original noisy data. (b) The result of the approach (Nsnet2) which has a missing word. (c) The result of the Putt model which gives the correct transcription result. (d) The clean speech which gives the correct STT result.
%     }
%     \label{Example}
% \end{figure}

In this study, we introduce a novel multi-stage speech enhancement model which is not based on diffusion models. The key distinction from previous approaches\cite{Wang,Lemercier} lies in our use of a supervised model instead of a stochastic diffusion model as the generative component. This design choice significantly reduces inference time. 
As will be demonstrated throughout this paper, our approach achieves superior performance compared to traditional one-step speech enhancement models.

We adopt terminology from golf, referring to our first speech enhancement program as Approach and the second-stage model as Putt. The multi-stage process composed of Approach and Putt differs from conventional methods in the following ways:

First, during Putt, we do not use generative models such as diffusion models but instead employ another supervised learning method. This allows for significantly faster computation compared to the time-consuming diffusion models.

Second, while the Approach network is trained to minimize the distance to clean speech, the Putt stage focuses on reducing artifacts. In this context, we define the magnitude of an artifact as the shortest distance to the straight line connecting the clean speech and noisy sound.

Artifacts in speech enhancement have been studied using the orthogonal projection method\cite{Iwamoto1,vincent}. This approach assumes distinct subspaces for speech and noise, considering any components outside these subspaces as unnatural artifacts. The speech and noise subspaces are spanned from the available data, and a projection matrix 
 is constructed accordingly. The artifact in the enhanced speech signal 
 was defined as the difference between the enhanced sound and the sound projected using the projection matrix\cite{Iwamoto1,vincent}.
 
When using the orthogonal projection method, a specific signal subspace is assumed. However, if this assumption does not match the actual environment, performance degradation may occur. This issue becomes more pronounced when the noise is non-stationary, as the projection may fail to capture the varying noise characteristics. Moreover, increasing the size of the projection matrix to better estimate the noise subspace can lead to excessive computational costs due to high-dimensional matrix operations. 
Instead of defining the artifact in the above manner, we adopted a new approach that defines it as the minimum distance to a natural sound that contains the same information. 

%Fig. \ref{Example} illustrates an example of the results produced by our model. The noise-mixed sound shown at the top (Fig. \ref{Example}(a)) demonstrates the significant addition of noise to a clean speech signal (Fig.  \ref{Example}(d)). When this signal is processed using a speech recognizer Whisper\cite{whisper}, it is incorrectly transcribed as 'and there's not for any other purpose' after applying the Approach (the first predictive model) to the noisy sound (Fig. \ref{Example}(b)), the noise is significantly reduced, as shown in Fig. \ref{Example}(b). However, the challenge lies in the presence of strong artifacts, which results in the wrong words. By utilizing our Putt model, we can restore the sound, as illustrated in Fig. \ref{Example}(c), leading to the correct recognition of 'It is not for any other purpose'

\section{Proposed Method}
\subsection{Artifact}
Let us represent a sound as a $T$-dimensional vector, which will be written here in boldface. 
Consider a sound $ {\bf X} = {\bf S + N}$,
 where
 ${\bf S}$, ${\bf N}$   $\in \mathbb{R}^{T}$ are speech signal and noise, respectively. The problem we aim to solve is how to extract the clean speech signal ${\bf S  }$ given a noisy sound {\bf X}. This problem will be addressed using a neural network based on deep learning.

The sounds we hear in our daily lives through our ears feel natural. This is because, even though multiple sound sources may be present, each source produces sounds corresponding to its own characteristics. In other words, we perceive sounds as natural when they result from a linear combination of multiple sources.
However, when a speech enhancement model excessively removes frequency components shared by both noise and speech signals, nonlinear interference between the noise and speech occurs, making the sound feel unnatural. Based on this principle, we define the artifact vector as follows.

{\it Definition- }  Let $\tilde{\bf X}$ be the output of the speech enhancement neural network for the input of the noisy data ${\bf X}$, and $\bf{S}$ be the corresponding clean speech data. 
Then we define the error vector $\boldsymbol{ \xi} \equiv \tilde{\bf X}-{\bf S}=\boldsymbol{ \xi_{\perp}}+\boldsymbol{\xi_{\parallel}}$. The artifact vector   is defined to be the perpendicular vector $\boldsymbol{ \xi}_{\perp}$  in this work ( See Fig.\ref{Schematic} ) and the parallel component vector $\boldsymbol{\xi}_{\parallel}$ is coined here as the proximity vector.

The artifact  ${\boldsymbol{\xi_\perp}}=\tilde{\bf X} - \tilde{\bf X}^{\prime} $ and
the proximity $\boldsymbol{\xi}_{\parallel}
=\tilde{\bf X}^{\prime} -{\bf S}$  can be rewritten as 
\begin{eqnarray}
   \boldsymbol{\xi}_{\perp} (\tilde{\bf X})  &=&
   (\tilde{\bf X}-{\bf X}) - \frac{{\bf S} -{\bf X}}{ |{\bf S} -{\bf X}|} \cdot (\tilde{\bf X}-{\bf X}) \frac{{\bf S} -{\bf X}}{ |{\bf S} -{\bf X}|} \\ 
   \boldsymbol{\xi}_{\parallel}(\tilde{\bf X}) 
   &=& \tilde{\bf X} - \boldsymbol{\xi}_{\perp} (\tilde{\bf X}) -{\bf S}
\end{eqnarray}
See Fig. \ref{Schematic} for the derivation of the expression.

\subsection{Loss functions}

We now aim to devise an algorithm to obtain a clean speech signal ${\bf S(X) }$ corresponding to any given noisy sound {\bf X}. 
Initially, the noisy sound {\bf X} is processed using a speech enhancement model {\bf Sp}({\bf X}) to produce the enhanced speech signal 
$\tilde{\bf X}$={\bf Sp}({\bf X}). We call this process the Approach. For the speech enhancement model used in the Approach stage, we adopted the same architecture as the Putt model—whose details will be described later—except that it does not include LSTM or dilated dense blocks, and the unpooling layer is implemented with a transposed convolution. 

When we train the network for the speech enhancement model 
${\bf Sp(X)}$, we use MSE loss function.
\begin{eqnarray}
\mathcal{L}_{approach}=\mathbb{E}_{\bf S , {\bf N }}|| {\bf Sp}({\bf X})-{\bf S(X)}||^{2}. 
\end{eqnarray}
%We may use various loss functions for better performance of the speech enhancement.

%As illustrated in Fig. \ref{Example}, while 
%{\bf Sp}({\bf X}) effectively reduces the noises, and it also removes some of the essential components of clean speech.
We introduce an additional  process which is intended to reduce the artifact $\xi_{\perp}$ using
 a vector function ( we call here a Putt function) 
${\bf \Xi}(~ \cdot~ ;{\bf X})$ to achieve a better quality output of the first Putt {\bf X}$_{putt}^{(1)}$;

\begin{eqnarray}
{\bf X}_{putt}^{(1)}({\bf \tilde{X}})=\tilde{{\bf X}} - \mathbf{\Xi}(\tilde{{\bf X}}; {\bf X}).
\label{xap}
\end{eqnarray}

Here, the original noisy sound ${\bf X}$ is concatenated to the input of the Putt function to keep its original clean speech. 
It will be optimal for the Putt function $\boldsymbol{\Xi}(\tilde{{\bf X}}; {\bf X})$ to be the artifact vector $\boldsymbol{\xi_{\perp}}(\tilde{{\bf X}};{\bf X})$, then the sound ${\bf X}_{putt}$ would be completely natural i.e. in the line of ${\bf S}$ and ${\bf X}$.
Thus, to train the neural network for the Putt function, we use 
the following loss function;
\begin{eqnarray}
\mathcal{L}_{putt}&=&\mathbb{E}_{\bf S, {\bf N }} || \boldsymbol{\xi_{\perp}}({\bf Sp}({\bf X});{\bf X},{\bf S}) \nonumber \\ && ~~~~~~~~~~~- \mathbf{\Xi}({\bf Sp}({\bf X}); {\bf X})  ||^{2}
\end{eqnarray}

\begin{figure}[htb]
 \centering
  \centerline{\includegraphics[width=\columnwidth ]{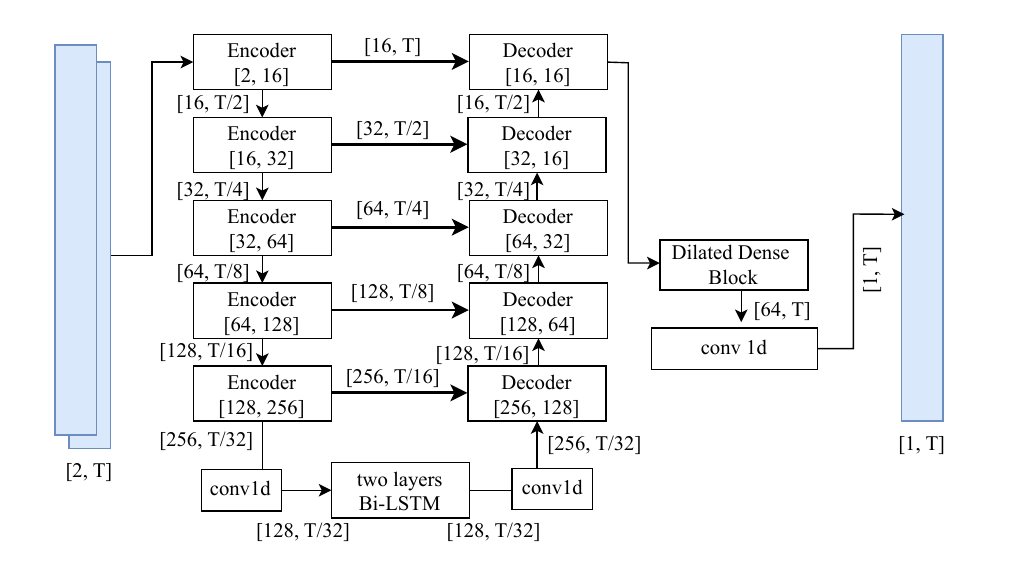}}

\caption{
 Illustration of the modules of the proposed network
for predicting artifacts in enhanced speech. The
time-domain artifact prediction network (the Putt model) is based on the
convolutional-recurrent-neural-network (CRN) and consists of a
time-domain U-Net combined with two layers of Bi-LSTM.  
}

\label{model_scheme}

\end{figure}

\subsection{Approach and Putt process}
A recursive or iterative application of these two stages — Approach → Putt → Approach → Putt — can be considered to progressively refine the enhanced output.
The $i$-th ($i$=2,3,...) speech enhancement and its Putt is given by
\begin{eqnarray}
   \tilde{{\bf X}}^{(i)}&=&{\bf Sp}({\mathbf X}_{putt}^{(i-1)}) \\
{\bf X}^{(i)}_{putt} &=&\tilde{{\bf X}}^{(i)}
- \mathbf{\Xi}({\tilde{{\bf X}}^{(i)}}; {\bf X}).
\end{eqnarray}

%\section{Experiments}

\subsection{Datasets}

    We use the VoiceBank-DEMAND dataset\cite{valentini-botinhao}, which consists of clean utterances from 28 speakers in the Voice Bank corpus\cite{veaux}, and noisy versions created by mixing them with two synthetic noises (babble and speech-shaped) and eight real-world noise recordings from the DEMAND database\cite{thiemann}. The training set includes mixtures at SNR levels of 0, 5, 10, and 15 dB, while the test set uses 2.5, 7.5, 12.5, and 17.5 dB\cite{valentini-botinhao}.

    All audio signals are originally sampled at 48 kHz and are resampled to 16 kHz for our model training and evaluation. For this resampling, we use the Python function ‘librosa.load’, which is commonly employed for audio processing.
    
    Finally, we prepare a total of 11,572 clean-noise pairs for training and 824 pairs for evaluation. During training, we apply a batch-wise shuffling strategy in which the clean signals are paired with randomly shuffled noise samples at each epoch. We use a batch size of 32, and each input segment consists of 8192 samples (0.5 seconds at 16 kHz). During training, we use the AdamW optimizer, which incorporates decoupled L2 regularization (PyTorch default weight decay = 0.01). The learning rate is set to 1e-5.

\begin{figure}[htb]
 \centering
  \centerline{\includegraphics[width=\columnwidth ]{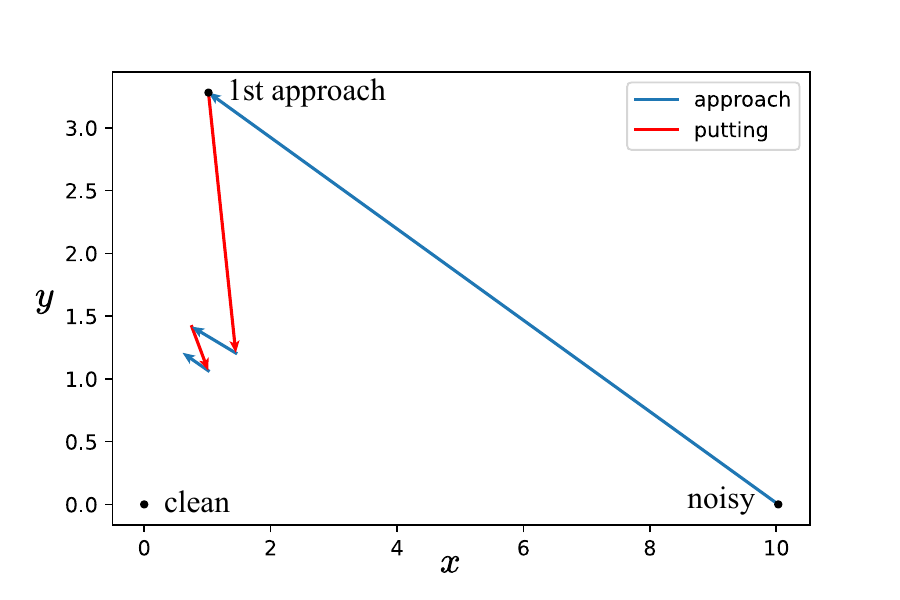}}

\caption{
The trajectory of waveform transformations in the 2D projection space defined by 
$x({\mathbf Z}) = ({{\mathbf Z}} - \mathbf{S}) \cdot 
   \hat{\boldsymbol{\xi}}_\parallel(\tilde{\mathbf X})
   , \quad y({\mathbf Z}) = ({\mathbf{Z}} - \mathbf{S}) \cdot 
   \hat{\boldsymbol{\xi}}_\perp (\tilde{\mathbf X})$,
where
$\hat{\boldsymbol{\xi}} =\frac{\boldsymbol{\xi}_\parallel}{|\boldsymbol{\xi}_\parallel|} $ and 
$\hat{\boldsymbol{\xi}}_\perp =\frac{\boldsymbol{\xi}_\perp}{|\boldsymbol{\xi}_\perp|} $. Roughly speaking, here 
x can be regarded as a component of the proximity vector that indicates how close it is to the clean signal, while 
y can be regarded as a component of the artifact vector. 
}

\label{nsnet_onsample_space}

\end{figure}

\subsection{The Putt Network}

 As illustrated Fig. \ref{model_scheme}, the architecture of the Putt network is based on the waveform-domain Convolutional-Recurrent-Network(CRN)\cite{crn1, crn2}, which is often used for speech enhancement in the time-domain. Our network consists of five encoders and five decoders with convolutional networks, arranged symmetrically. Each encoder and decoder layer is connected by skip-connections. 
 The input tensor of the network is a concatenation of two speech waveform vectors: {\bf Sp}({\bf X})($\in$ $\mathbb{R}^{T}$; the enhanced speech sound) and ${\bf X}$($\in$ $\mathbb{R}^{T}$; the noisy speech sound), where the T represents the length of each speech sound. In Fig. \ref{model_scheme}, [ , ] represents [the number of channels, T]. The output tensor of the network is a single channel containing one speech waveform, denoted as $\mathbf{\Xi}$ ($\in$ $\mathbb{R}^{T}$), which predicts the artifact vector $\boldsymbol{\xi}$. 
  Additionally, our network features a bottleneck structure with two layers of Bi-LSTM as a recurrent network between the encoders and decoders. The encoder and decoder blocks are composed of a combination of two CBP (convolution layers, batch normalization, and PReLU).
 % as shown in Fig. \ref{encdec}. 

% \begin{figure}[htb]
%  \centering
%   \centerline{\includegraphics[width=\columnwidth ]{figs/encdec.pdf}}

% \caption{
% The encoder and decoder of the network. Each encoder and decoder
% have four CBP (convolution + batch normalization + PReLU) layers and delated dense block.
% }

% \label{encdec}

% \end{figure}
Due to the local nature of convolution operations in the waveform domain, our model exhibits limitations in modeling low-frequency components that demand long-range dependencies. To mitigate this, we incorporate three dilated dense blocks\cite{Pandey,Huang} before the pooling/unpooling stages, allowing the network to capture a wider temporal context.

Pooling and unpooling modules are inserted between the encoder and decoder. Both modules employ a CBP (Conv–BN–PReLU) structure; however, in the unpooling stage, the standard convolution layer is replaced with a sub-pixel convolution to effectively reconstruct the temporal resolution\cite{Pandey}.

\begin{figure}[htb]
 \centering
  \centerline{\includegraphics[width=\columnwidth ]{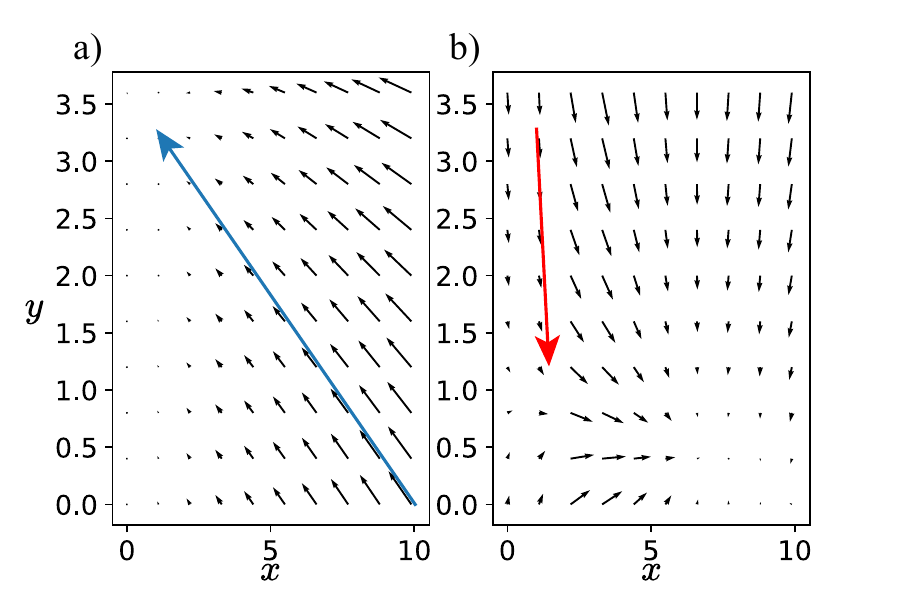}}

\caption{ a) The 2D projection of the vector field ${\bf F}_{approach}$, illustrating the direction in which the system is driven by the Approach mechanism. The blue arrows indicate the actual trajectory of the system under the influence of Approach.
b) The vector field induced by Putt, showing the direction in which the system is driven. The red arrows represent the actual trajectory of the system under the influence of Putt.}
\label{vectorfield}
\end{figure}

\section{Results}

As shown in Fig. \ref{nsnet_onsample_space}, by repeatedly applying the Approach and Putt procedures, the resulting signal progressively approximates the clean speech.
  In this figure, we defined a projected two-dimensional space $ (x({\bf{Z}}),y({\bf{Z}}))$ for high dimensional vector ${\bf Z}\in\mathbb{R}^T$.

%If a speech enhancement model were perfect—i.e., capable of generating completely clean speech—then any method, whether single-stage or multistage, would yield the same result. However, in practice, neural network training is inherently imperfect due to limitations in training data and model capacity. Consequently, a non-ideal enhancement model does not produce a truly clean signal but rather a signal that deviates in some way from the clean target.

To better understand how our multi-stage process leads to the improvement of the speech enhancement, we visualize the model using the following procedure. We first select an audio sample ${\bf X}$ (p301\_116 speech from the VoiceBank-Demand 56 speakers trainset\cite{valentini-botinhao}) that contains both speech and noise. Then, we consider a two-dimensional space spanned by two functions, $\boldsymbol{\xi}_{\parallel}(\tilde{\bf X} )$ and
$\boldsymbol{\xi}_{\perp}(\tilde{{\bf X}} )$, where $\tilde{\bf X}$
denotes the enhanced version of ${\bf X}$.
For any point ${\bf Z}$ in this two-dimensional space, let 
$\tilde{\bf Z}$ denote the output obtained by applying the speech enhancement model to ${\bf Z}$. We then define a vector field ${\bf F}_{approach}$ over this space based on the behavior of the model at each point.

\begin{eqnarray}
   {\bf F}_{approach}({\bf Z}) = \left((\tilde{{\mathbf Z}} - \mathbf{Z}) \cdot 
   \hat{\boldsymbol{\xi}}_\parallel (\tilde{\mathbf X})
   ,  (\tilde{\mathbf{Z}} - \mathbf{Z}) \cdot 
   \hat{\boldsymbol{\xi}}_\perp(\tilde{\mathbf X})\right)
\end{eqnarray}

The Approach field $\mathbf{F}_{approach}$ describes the tendency of the Approach speech enhancement model. As shown in the left panel of Fig. \ref{vectorfield} a), ${\bf F}_{approach}$ does not converge toward the clean speech point, but rather converges toward unintended regions. This phenomenon arises from the imperfection of the speech enhancement model, and it demonstrates that following the vector field induced by the model does not lead to the recovery of clean speech. The blue arrow denotes the actual process where the sound is changed by the Approach.

% As shown in the figure, the magnitude of the vector gradually decreases as it proceeds in the direction of the vector, eventually coming to a stop. However, the stopping point is not the point of the clean speech.
Fig. \ref{vectorfield} b) illustrates the general tendency of the Putt network's vector field, indicating the direction toward which each point is guided. Notably, the vectors predominantly point downward, corresponding to a decrease in the 
y-value. As shown by the red line, the trajectory moves downward, implying a substantial reduction in the artifact component.

When only the Approach is applied, the process halts at the point where the Approach field vanishes. At that point, the Putt acts to escape from the vanishing field region. From there, further improvement is achieved through the Approach again. The Putt serves to reduce artifacts, enabling the next Approach to progress further.
Even when the Approach and Putt processes are repeated infinitely, the output does not perfectly converge to the clean signal; however, substantial improvement is observed.

\begin{figure}[htb]
 \centering
  \centerline{\includegraphics[width=\columnwidth ]{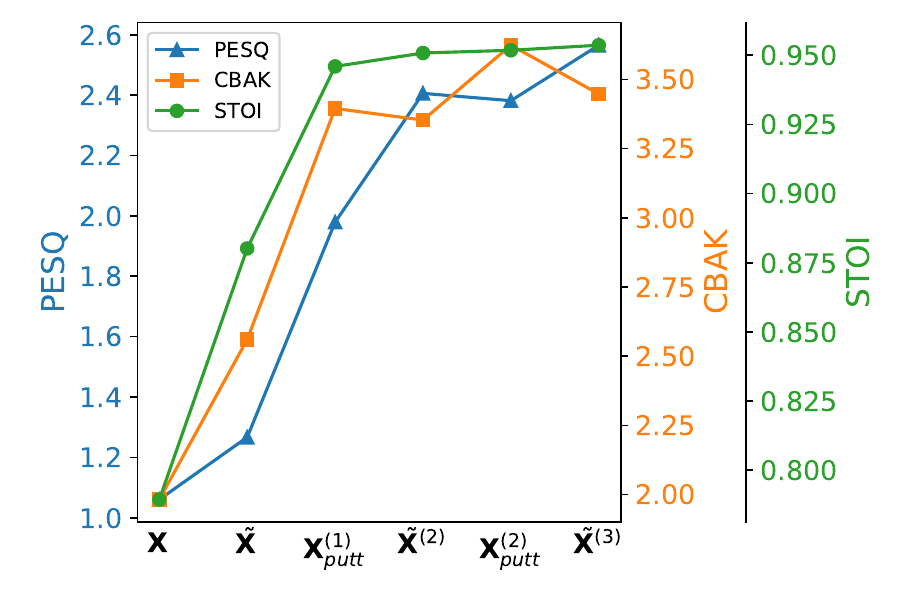}}
\caption{PESQ, CBAK, and STOI scores for a noisy speech signal processed alternately with the Approach and Putt models. $\bf X$ is the original noisy signal. $\tilde{\bf X}$ is the first speech enhancement of $\bf X$. $\tilde{\bf X}^{(i)}$ is the result of the Approach of ${\bf{X}}^{(i-1)}_{putt}$. ${\bf{X}}^{(i)}_{putt}$ is the output of the Putt of $\tilde{\bf{X}}^{(i)}$}

\label{nsnet_onesample_metric}

\end{figure}

\begin{figure}[htb]
 \centering
  \centerline{\includegraphics[width=\columnwidth ]{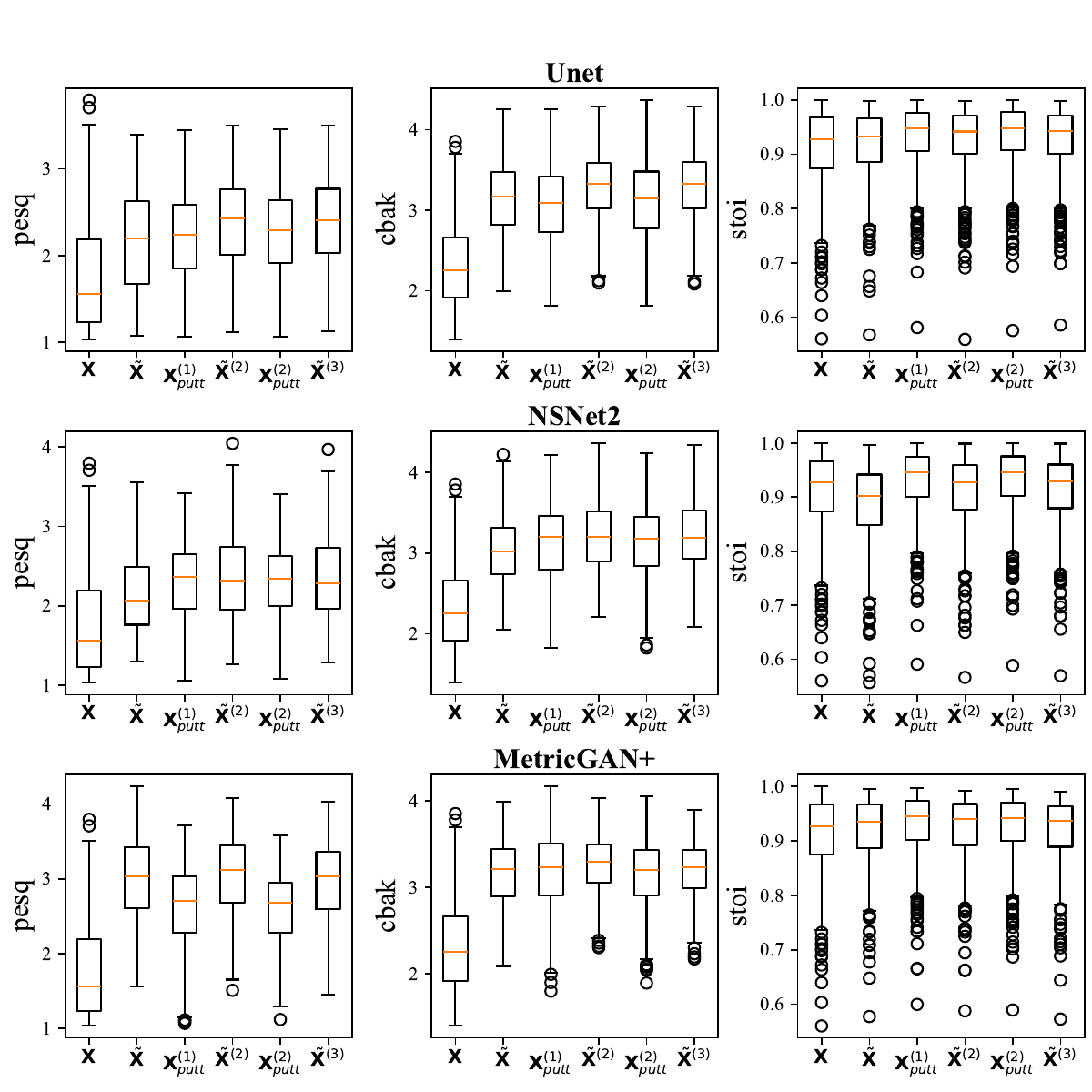}}

\caption{The results of applying Approach and Putt using a UNet-based SE model(which adopts the same encoder/decoder architecture as the Putt network, but excludes delated dense block and LSTM layer) trained with Voicebank-demand dataset on wave-domain signal, NSNet2\cite{BRAUN},  and MetricGAN+\cite{FU}. This figure illustrates the PESQ, CBAK, and STOI results for the samples from the VoiceBank-Demand test dataset. The evaluations were conducted at three specific SNR levels: 2.5, 7.5, and 12.5 dB.}
\label{metrics_10percent}

\end{figure}

Fig. \ref{nsnet_onesample_metric}  illustrates that applying Approach and Putt alternately to a single speech file containing one sentence leads to improvements in perceptual quality metrics such as PESQ\cite{pesq}, CBAK\cite{cbak}, and STOI\cite{stoi}. While the scores do not increase indefinitely, the enhancements are substantial enough to indicate meaningful improvements in sound quality. This phenomenon is not limited to a single utterance: similar trends are consistently observed when aggregating results over a large number of speech files or when replacing the Approach module with different speech enhancement models (See Fig. \ref{metrics_10percent}).

%PESQ (Perceptual Evaluation of Speech Quality) is an objective metric standardized by ITU-T (P.862) for assessing the perceptual quality of speech signals. It compares the original clean signal with the processed (possibly distorted) signal using a perceptual model that simulates the human auditory system. 
%CBAK (Background Intrusiveness in the Composite Measures) is part of the ITU-T P.835 standard, which evaluates the perceived intrusiveness of background noise in a processed speech signal. A higher CBAK score indicates that the background noise is perceived as less intrusive, reflecting more effective noise suppression.
%STOI (Short-Time Objective Intelligibility) is a metric designed to predict the intelligibility of speech, particularly in noisy conditions. It computes the correlation between the temporal envelopes of the clean and processed speech signals over short-time segments. STOI values range from 0 to 1, with higher scores indicating greater intelligibility.

\section{CONCLUSION}

We show that alternating between our proposed Putt model and a standard enhancement model outperforms using either alone. This method is potentially applicable beyond speech, including image and other noise reduction tasks. Unlike conventional models that aim for clean data, Putt directs outputs toward the foot of the perpendicular from the noisy point to the line connecting clean and noisy data. As noted, this line consists of natural, artifact-free sounds.

In the 2-D projected space, we visualized the vector fields of Approach and Putt to observe the directional tendencies each model applies to the signal. The illustrations reveal that signals residing in regions with vanishing vectors—where a single model can no longer make progress—are moved by the other model, thereby enabling the multi-stage method to achieve superior results.

% \section{acknowledgments}

{}

\end{document}